\def\year{2022}\relax
\documentclass[letterpaper]{article} 
\usepackage{aaai22}  
\usepackage{times}  
\usepackage{helvet}  
\usepackage{courier}  
\usepackage[hyphens]{url}  
\usepackage{graphicx} 
\usepackage{natbib}  
\usepackage{caption} 
\DeclareCaptionStyle{ruled}{labelfont=normalfont,labelsep=colon,strut=off} 
\usepackage{booktabs}

\usepackage{amsfonts}       
\usepackage{nicefrac}       
\usepackage{microtype}      
\usepackage{lipsum}
\usepackage{amsmath}
\usepackage{multirow}
\usepackage{amssymb}
\usepackage{subcaption}
\usepackage{xr}
\usepackage{xcolor}
\usepackage{footmisc}
\usepackage[linesnumbered,ruled,vlined]{algorithm2e}
\title{Correcting Sociodemographic Selection Biases for Population Prediction from Social Media}
\author {
    Salvatore Giorgi,\textsuperscript{\rm 1}
    Veronica E. Lynn,\textsuperscript{\rm 2}
    Keshav Gupta,\textsuperscript{\rm 2}
    Farhan Ahmed,\textsuperscript{\rm 2}\\
    Sandra Matz,\textsuperscript{\rm 3}
    Lyle H. Ungar,\textsuperscript{\rm 1}
    H. Andrew Schwartz\textsuperscript{\rm 2}\\
}
\affiliations {
    \textsuperscript{\rm 1} University of Pennsylvania \\
    \textsuperscript{\rm 2} Stony Brook University \\
    \textsuperscript{\rm 3} Columbia University \\
    sgiorgi@sas.upenn.edu, has@cs.stonybrook.edu
}

\newcommand{\name}[0]{\textsc{Robust Poststratification}\xspace}

\begin{document}

\maketitle

\begin{abstract}

Social media is increasingly used for large-scale population predictions, such as estimating community health statistics. 
However, social media users are not typically a representative sample of the intended population --- a ``selection bias''.
Within the social sciences, such a bias is typically addressed with \textit{restratification} techniques, where  observations are reweighted according to how under- or over-sampled their socio-demographic groups are. 
Yet, restratifaction is rarely evaluated for improving prediction.

\hspace{1em} In this two-part study, we first evaluate standard, ``out-of-the-box" restratification techniques, finding they provide no improvement and often even degraded prediction accuracies across four tasks of esimating U.S. county population health statistics from Twitter.
The core reasons for degraded performance seem to be tied to their reliance on either sparse or shrunken estimates of each population's socio-demographics.  
In the second part of our study, we develop and evaluate \name, which consists of three methods to address these problems: (1) \textit{estimator redistribution} to account for shrinking, as well as (2) \textit{adaptive binning} and (3) \textit{informed smoothing} to handle sparse socio-demographic estimates. 
We show that each of these methods leads to significant improvement in prediction accuracies over the standard restratification approaches.
Taken together, \name enables state-of-the-art prediction accuracies, yielding a 53.0\% increase in variance explained ($R^2$) in the case of surveyed life satisfaction, and a 17.8\% average increase across all tasks.

\end{abstract}

\section{Introduction}
\label{section:introduction}
Digital language has shown promise for inexpensive large-scale population measurement~\cite{coppersmith2015adhd,mowery2016towards}. 
Twitter, for example, has been used to track public opinion~\cite{o2010tweets, miranda2015twitter} and measure community health~\cite{mowery2016towards,abebegiorgi2020quantifying,de2013social}.
The passive assessment of community characteristics that are otherwise expensive to obtain offers tremendous opportunities for both researchers and practitioners, but it also poses a challenge that is often overlooked: predictions made from social media are often prone to significant bias resulting from non-representative samples.

Although the user bases of social media platforms are diversifying, they do not accurately reflect the general population \cite{pewresearch,greenwood2016social}. 
For example, Twitter users typically are younger and have a higher median income \cite{pewresearch}.  
Per location, such as counties in the U.S., these biases can further differ. 
As a result, samples collected from Twitter are not representative of the populations they are intended to model, leading to a  ``selection bias" that can potentially skew results. 

In this study, we address the issue of selection bias when using spatially aggregated Twitter language to measure community health and well-being. 
We estimate age, gender, income, and education distributions of a geolocated Twitter sample through pretrained socio-demographic models.  
When compared to known distributions of the community (via the U.S. Census), these inferred socio-demographic variables allow us to quantify the selection bias per observation (a county in this case). 
Using these insights, we estimate county-level language features, weighting each county member according to how over- or under-represented they are within the community's known socio-demographic distribution.

While addressing selection bias is a common procedure in many quantitative social sciences, it is primarily used to improve in-sample correlational statistics \cite{berk1982selection,winship1992models}. 
In contrast, attempts to address selection bias to improve predictive models (i.e. supervised NLP) are rare. 
One potential explanation for this gap is that socio-demographic information is rarely available in predictive contexts, which rely on observable data rather then self-report questionnaires. 
However, recent work demonstrates that estimates of demographics from language can be highly associated with self-reported demographics~\cite{zhang2016your,chen2015comparative}, which could provide a means to estimating and correcting demographic selection bias. 

Similar to in-sample corrections of selection biases, one would expect that the accuracy of predictive models (e.g., predictions of \textbf{representative} health outcomes) can be improved by taking account of observable selection biases. 
Here, we show that this is not the case: applying standard in-sample solutions (e.g., post-stratification and raking) leads to a decrease in performance when predicting representative county health. 
Upon investigation, we identify that this drop in performance arises from two problems: (1) the use of estimated socio-demographics and (2) the sparse socio-demographics bins (i.e., socio-demographic subgroups which are sparsely populated in a sample when compared to known populations). 
Building on these findings, we propose novel solutions to each of these problems in the form of (1) estimator redistribution and (2) adaptive binning and informed smoothing. 

This paper presents methods and results in two stages. 
First, we define and contextualize the problem of selection bias, presenting existing methods (\textbf{Out-of-the-box Correction Techniques}) and highlighting the fact that standard methods fall short in our setup (\textbf{Results based on Existing Methods}). 
Second, we introduce novel methods to handle challenges common to selection bias correction in \textbf{Improving Correction Techniques} and present results in \textbf{Results: Estimation and Sparsity Challenges}.

\paragraph{Contributions} 
Our key contributions include:
  (1)  Introduce the problem of selection bias correction for supervised NLP, including standard methods from other fields;
  (2) Show that standard reweighting techniques used widely in other fields often lead to degraded performance in predicting health statistics from social media language;
  (3) Identify the problems of standard reweighting and develop methods to mitigate them;
  (4) Apply these techniques as \name to obtain state-of-the-art prediction accuracies of county life satisfaction and health.
  We also open-source all code.\footnote{\label{code}Code: \url{https://github.com/wwbp/robust-poststratification}} and data\footnote{\label{supp_and_data}Data and Supplemental Materials: \url{https://osf.io/ae5w6}}


\section{Problem Statement: Selection Bias Correction\protect\footnote{See Table \ref{table:glossary} for definitions of all terms.}}
Given hierarchical data where lower level individual data points (i.e., Twitter users) are nested within a population (i.e., U.S. county), we wish to estimate the 
representative population-level expectation, $\mu_{X_i}$, from the lower level data. 
For example, when correcting for selection bias of language on Twitter, $X$ is a vector of linguistic features for which we wish to derive a representative mean. 


Simple averaging methods fail to account for differences between the observed sample (individuals in our Twitter data mapped to a county) and the target population (the entire population of a county) for whom a measurement is desired. 
Thus, with respect to a target population, the measurements over the sample are biased, i.e. suffer a selection bias. 
More formally, we define $d=\lbrace d_m\rbrace$ to be a set of individual level auxiliary variables (in our case, $d=\lbrace$age, gender, income, education$\rbrace$), $\text{Q}_i(d)$ to be the distribution of our sample (those for whom we have a measurement in our data set) and $\text{P}_i(d)$ to be the distribution of the target population (those for whom a measurement is desired) in U.S. county $i$. 
Then, following \citet{shah2020predictive}, we take selection bias to mean that the sample distribution is dissimilar from a theoretically-desired distribution (the census-measured population distribution in this case):
$$\text{P}_i(d) \nsim \text{Q}_i(d).$$
We can then view selection bias correction as estimating a correction factor for the given set of auxiliary variables $d$: 
\begin{equation}
    \psi_i(d) = \frac{\text{P}_i(d)}{\text{Q}_i(d)},
    \label{eq:weight}
\end{equation}
\noindent such that our goal of estimating $\mu_{X_i}$, the population expectation of the individuals' features $X_i$ for community $i$, can be written as
\begin{equation}
    \hat{\mu}_{X_i} = \frac{1}{N_i}\sum_{j\in U_i} \psi_i(d)r_j(x_j).
    \label{eq:pop average}
\end{equation}
Here $U_i$ is the set of individuals in community $i$, with $N_i = \big|U_i\big|$, and $r_j$ is some kernel function. 
Note that Eq. \ref{eq:weight} is similar to the Kullback-Leibler divergence \cite{kullback1951information}. 
Since we would like a multiplicative correction factor, we do not take the log of the ratio.

\begin{table}[t!]\centering
\fontsize{9}{9}\selectfont
\begin{tabular}{p{0.1\columnwidth} | p{0.8\columnwidth}}
\toprule
 & Definition \\ \hline
$d^{(s)}$ & Twitter user's predicted socio-demographics value in the Twitter sample $s$\\
$d^{(t)}$ & Twitter user's redistributed socio-demographics value in the target distribution $t$\\
$d_m$ & Twitter user's predicted value for socio-demographic $m$ (age, gender, income, education)\\
$D_{d_m}$ & Partition of the socio-demographic $d_m$  \\
$D^{(h)}_{d_m}$ & Subset of partition $D_{d_m}$ \\
$i$ & Index, population level observations (U.S. counties) \\
$j$ & Index, individual level observations (Twitter users) \\
$m$ & Index, socio-demographics (age, gender, income, education) \\
$h$ & Index, socio-demographic partitions \\
$d$ & Set of socio-demographic variables (subsets of $\lbrace$age, gender, income, education$\rbrace$) \\
$k$ & Smoothing parameter  \\ 
$\min^{(s)}_h$ & Minimum socio-demographic value in subset $D^{(h)}_{d_m}$ in our Twitter sample $s$  \\
$\max^{(s)}_h$ & Maximum socio-demographic value in subset $D^{(h)}_{d_m}$ in our Twitter sample $s$  \\
$\min^{(t)}_h$ & Minimum socio-demographic value in subset $D^{(h)}_{d_m}$ in our target distribution $t$  \\
$\max^{(t)}_h$ & Maximum socio-demographic value in subset $D^{(h)}_{d_m}$ in our target distribution $t$  \\
$N_i$ & Cardinality of $U_i$ (number of Twitter users in county $i$) \\ 
$\text{P}_i(d)$ & Distribution of the target population (U.S. Census) \\
$\text{Q}_i(d)$ & Distribution of our sample (Twitter sample) \\
$s$ & Sample distribution (Twitter users) \\
$t$ & Target distribution (U.S. Census) \\
$U_i$ & Set of all individuals within a population (Twitter users county $i$) \\ 
$X_i$ & Population level observation (county level linguistic feature) \\
$x_{i,j}$ & Individual level observation (Twitter user linguistic feature) \\
$\psi_i(d)$ & Correction factor  \\ 
$\mu_{X_i}$ & County level expectation of $X_i$ \\ 
\bottomrule
\end{tabular}
\caption{Definitions for notation used throughout the paper.}
\label{table:glossary}
\end{table}

This formulation, rooted in the literature on reweighting and post-stratification techniques from economics and social science~\cite{kalton2003weighting,hoover2018big}, includes several useful abstractions. 
First, $d$ can contain any number of auxiliary variables, though, here $d$ is a set of socio-demographics of the Twitter users (e.g., various combinations of age, gender, income, and education). 
Higher level populations need not be limited to counties (e.g., cities and countries), nor do these populations need not be limited to spatial regions.
Further, although our focus is social media-based community measurements, this formulation could be used for other types of data: estimating consumer metrics per household~\cite{alexander1987class}, state polling~\cite{park2004bayesian}, and, generally, individual-level data from a biased sample of a population. 


\subsection*{Bias in Social Media Population Measurement}
\label{section:background}


Samples collected from Twitter aren't generally representative of the real-world populations that they are intended to model~\cite{mislove2011understanding,culotta2014reducing}. 
To some extent, this is attributable to unbalanced user demographics -- users skew young, toward one gender or the other, toward wealthy or poor, and toward urban rather than rural~\cite{pewresearch,greenwood2016social,hecht2014tale}.
Beyond demographics of who ``selects'' to use social media, data collection methods further contribute to selection biases.
The geotagging process can select certain ages and genders~\citep{pavalanathan2015confounds}, and races may be partially excluded due to language filters, prone to errors on region- or race-specific dialects such as African-American English~\cite{blodgett2016demographic}.


Limited work has been done to correct for selection biases on social media. 
Recently, \citet{wang2019demographic} presented a method for selection bias correction to create national population estimates from social media.  
They showed that one could use inferred demographics with a traditional post-stratification technique to produce more representative population statistics. 
We also use estimated demographics, but we find these traditional post-stratification techniques have problems which lead to degraded performance for predictive modeling. 
Other fields have presented social media-specific frameworks ~\citep{zagheni2015demographic} but without predictive evaluations.  


While population studies often attempt to correct for selection bias, few have explored the use of corrections to improve predictive modeling. 
Non-representative samples can have a significant impact on model performance. 
For example, \citet{weeg2015using} used mentions of diseases on Twitter and nearly doubled predicting prevalence rates for 22 diseases after limiting analysis to disease prevalence amongst known Twitter users. 
Using Twitter to predict elections, \citet{miranda2015twitter} explored selection bias as a reason for inconsistent election predictions. 
Attempting to construct stratified samples, they concluded that results were encouraging but lacked sufficient data to make predictions. 
Our method works even in cases such as this where traditional restratification isn't feasible. 

Closest to our work, we build on ideas from \citet{culotta2014reducing} who explored reweighting schemes for predicting county-level health statistics. 
They reweighted instances according to users' predicted gender and race, leading to improved predictions for 20 out of 27 variables. 
However, their evaluation was limited to the top 100 most populous counties, which are primarily homogeneous urban centers. 
In contrast, our work explores methods for cases where the data is not homogeneous and/or when data is sparse. 
Further, we provide a more comprehensive evaluation and correct for more variables (e.g. age, gender, education, and income).

\section{Data\footref{supp_and_data}}
\label{section:data}

Our training data is broken into two pieces: (1) the biased sample from tweets and (2) representative population-level survey percentages. The biased sample consists of Twitter data which we want to aggregate to the community level in such a way that its socio-demographic makeup matches that of our representative population, the U.S. Census. 

\paragraph{Biased Sample: Twitter} We use the open source County Tweet Lexical Bank --- a U.S. county-mapped Twitter data set built over 1.6 billion tweets~\cite{giorgi2018remarkable}. 
The County Tweet Lexical Bank Twitter data was pulled from July 2009 to February 2015, geolocated to U.S. counties (i.e., county FIPS codes, which are unique numeric identifiers for U.S. counties) via self-reported location information in public account profiles and latitude / longitude coordinates (see \citet{schwartz2013characterizing} for the county mapping/geolocation process) and then filtered to contain only English tweets~\cite{langid}.
At a high level, the county mapping process is as follows: (1) if a tweet object contains latitude / longitude coordinates, we can trivially map that tweet to a county, (2) if a Twitter account has self-reported location information in the profile (e.g., ```living in NYC") we match the location string in the profile to U.S. cities which can be mapped to counties. 
The data was then limited to Twitter accounts with at least 30 posts and U.S. counties represented by at least 100 such unique accounts. 
The final Twitter data set consists of 2,041 U.S. counties with 6.06 million users.

\paragraph{Representative Population: U.S. Census} Five year estimates (2011-2015) for age, gender, education, and income were obtained from United States Census Bureau's 2015 American Community Survey (ACS).
Age records contain the percentages of people within age ranges 18-19, 20-24, 25-29, 30-34, 35-39, 40-44, 46-49, 50-54, 55-59, 60-64, and above 65.
The gender records consist of the percentages of males and females for each county. 
Percentages of income for the following bins: less than \$10,000, \$10,000-\$14,999, \$15,000-\$24,999, \$25,000-\$34,999, \$35,000-\$49,999, \$50,000-\$74,999, \$75,000-\$99,999, \$100,000-\$149,999, \$150,000-\$199,999, and greater than \$200,000.
Education is divided into two groups: percentage of the population with less than a Bachelor's degree and percentage higher than that of Bachelors. 

\paragraph{Outcomes} Selection bias correction is evaluated across four different community level out-of-sample prediction tasks, where we cross sectionally predict county level health variables (i.e., the outcomes or dependent variable in each task) from county level language on Twitter (i.e., the independent variables). 
The four tasks include two measures of objective health (heart disease and suicide mortality rates) and two measures of subjective health and well-being (Life Satisfaction and Poor or Fair Health). 
From the Centers for Disease Control and Prevention (CDC) we collected age-adjusted mortality rates for heart disease ($N=2,038$) and suicide ($N=1,672$), averaged across 2010-2015.
Life satisfaction scores are calculated as average individual level response to the question ``In general, how satisfied are you with in your life?'' (1 = very dissatisfied and 5 = very satisfied), averaged across 2009 and 2010 ($N=1,951$)~\cite{lawless2011predictors}. 
Finally, Poor or Fair Health was obtained from the County Health Rankings and is sourced from the Behavioral Risk Factor Surveillance System (BRFSS)~\cite{remington2015county}.
This is an age-adjusted measure of the percentage of adults who consider themselves to be in poor or fair health (i.e., percentage of adults who answered fair or poor to the question: ``In general, would you say that in general your health is Excellent/Very good/Good/Fair/Poor?"; $N=1,931$).
All $N$ reported above are a subset of the 2,041 counties with sufficient Twitter data, and are therefore the final number of observations in our four county tasks.

\paragraph{PEW: National Social Media and Twitter Use}
Since Twitter socio-demographic populations are not known at the county level we use National statistics collected from PEW's Social Media update~\cite{greenwood2016social}, including age, gender, income, and education for the years 2013-2016. 
Demographics were binned as follows: age (18-29, 30-49, 50-64, and 65+), gender (female / male), income (less than \$30,000, \$30,000-\$49,999, \$50,000-\$74,999, and \$75,000+; yearly), and education (high school grad or less, some college and college+). 
For each demographic bin we collect the percentage of the population who use social media and the percentage of the population who use Twitter. 
These percentages are averaged over the four years available, the closest available timeline to the Twitter sample. 
Using the total U.S. population we then calculate the percentage of people in each socio-demographic bin (i.e., total U.S. population $\times$ average bin percentage of people who use social media $\times$ average bin percentage of people on Twitter).

\paragraph{Ethics Statement} 
This study was reviewed and approved by an academic institutional review board, found to be exempt, non-human subjects data, with \textit{none to minimal} chance of harm to individuals.  
All raw data used in this study are publicly available.  
Our aggregate anonymized (manually checked for identifying information) language features by county are publicly available.\footref{supp_and_data}
For additional privacy protection, \textbf{no} individual-level estimates, intermediate information derived within the approach, will be made available.
The original tweets, which are publicly available, are not able to be redistributed due to Twitter's Terms of Service.


\section{Estimating Socio-demographic\\Bias from Language}

Sample socio-demographics are necessary in order to quantify and correct non-representation, but such information is not typically available in social media. 
We thus turn to socio-demographic estimates of our sample from their language. 
Such estimates have been validated in a number of contexts including Twitter~\cite{developing2014emnlp,matz2019predicting},\footnote{While perfection is not necessary to achieve benefit, excessive error would presumably prevent our approach from improving county-level predictions.} and a similar approach was used by~\citet{wang2019demographic}. 
We produced language-based estimates for four socio-demographic variables, which we will correct for selection bias: age, gender, income, and education. 
All four estimators are described below.
The median (or percentage) county values for our sample estimates versus census population statistics are given in Table \ref{table:county means}. On average, our Twitter sample appears younger and more educated than the population as a whole, but it is important to remember bias may differ from one county to the next and correction attempts to make each county more representative of its population. We note that gender is fairly evenly split across U.S. counties across all three categories: U.S. Census, Twitter sample, and PEW.

\begin{table}[tb!]\centering\fontsize{9}{9}\selectfont
\begin{tabular}{lcccc}\toprule
 & Age & \begin{tabular}[c]{@{}c@{}}Perc. \\ Female\end{tabular} & Income & \begin{tabular}[c]{@{}c@{}}Perc. Bachelor's\\  Degree\end{tabular}  \\\hline
Census & 39.3 & 50.4 & \$48,280 & 22.3 \\
Twitter & 22.1 & 53.8 & \$36,437 & 40.5 \\ \hline
PEW & 28.8$^*$ & 48.3 & \$58,660$^*$ & 42.9 \\ \bottomrule
\end{tabular}
\caption{County mean of medians or percentages. $^*$Imputed from bin percentages as median not provided by Pew. }
\label{table:county means}
\end{table}

We utilized estimated age, gender, income, and education, based on tweet language, using the following models.\footnote{Age, gender, and income estimation models were previously published, while education is novel to this paper.} 

\textbf{Age and Gender.} Age and gender estimates were based on a demographic predictive lexica~\cite{developing2014emnlp}. 
Sap inferred these models over a set of annotated users with self-reported age and gender (binary as multi-class gender was not available at the time) from Facebook, Twitter, and blogs. 
Accuracy of the estimates for age correlated with self-reported age at Pearson r = 0.86 and the estimated gender with an accuracy = 0.90 with self-reported gender. 
The model produced real values for age which were thresholded to between 13 and 80. 
For gender, output was a continuous score from negative (more male) to positive (more female). 
Because county statistics were limited to binary gender these were converted to 1 for ``female'' and 0 otherwise. 

\textbf{Income.} Income was estimated using the model built in \citet{matz2019predicting}. They collected a sample of 2,623 participants from Qualtrics in 2015 who reported their annual income and shared social media language. This model achieved an out-of-sample accuracy of Pearson r = .41 for estimated income as compared to true income. 
The model takes ngram frequencies as well as social media topic loadings from \citet{schwartz2013personality} as input. 

\textbf{Education.} 
An education classification model was built over a sample of users recruited from Qualtrics~\cite{preoctiuc2017beyond}. 
A total of 4,062 users reported education level and shared their Facebook status data. 
Mirroring Matz's income model, for each user, we extracted ngrams of length 1 to 3 and loadings for a set of 2,000 social media-based LDA topics~\cite{schwartz2013personality}. 
We used a multi-class linear-svc classifier and train on the following classes: (1) less than high school diplomam (2) high school diploma or Associate's degree, and (3) Bachelor's degree or higher. 
This model obtained an accuracy of .62 and an F1 score of .53 using 10-fold cross validation. 
We then used this model to predict class probabilities for Twitter accounts and, because most county data only indicated higher education percentages, collapsed the first two classes into a single class. 
This resulted in two final education classes: (1) less than a Bachelor's degree and (2) Bachelor's degree or higher.

\section{Study 1: Out-of-the-box Correction Techniques}
\label{section:methods_selection_bias}
Our approach to applying standard selection bias correction relies on three steps: (1) estimating socio-demographics, (2) creating weight factors, and (3) reweighting user level features and aggregating to the county (i.e., applying weight factors). 

In practice $\text{P}_i(d)$ and $\text{Q}_i(d)$ are unknown and must be estimated, typically by creating a partition $D_{d_m}$ of each socio-demographic variable $d_m$ into non-overlapping subsets $D^{(h)}_{d_m}$ where $\bigcup_h D^{(h)}_{d_m} = D_{d_m}$:
\vspace*{-4pt}
\begin{equation}
    \hat{\psi}_i(d) = \frac{\text{P}_i(d|d_m\in D^{(h)}_{d_m}, \forall m)}{\text{Q}_i(d|d_m\in D^{(h)}_{d_m}, \forall m)}.
    \label{eq:weight bins}
\end{equation}
Furthermore, the population distribution $P_i(d)$ is estimated using population percentages from known national surveys, in our case, the U.S. Census, and the sample distribution $Q_i(d)$ is estimated from our sample percentages:
\vspace*{-4pt}
\begin{equation}
    \hat{\psi}_i(d) = \frac{\text{perc}_{\text{pop}}(d|d_m\in D^{(h)}_{d_m}, \forall m)}{\text{perc}_{\text{samp}}(d|d_m\in D^{(h)}_{d_m}, \forall m)},
    \label{eq:weight perc}
\end{equation}
where $\text{perc}_{\text{pop}}$ and $\text{perc}_{\text{samp}}$ are the population and sample percentages, respectively. The non-overlapping subsets $D^{(h)}_{d_m}$ are referred to as bins throughout.

\subsection{Existing Methods}
We investigate two common methods for creating weight factors: (1) naive post-stratification and (2) raking, both of which are a form of post-stratification.
These two methods can be viewed as different ways of estimating the joint probability distribution in the population domain of the given socio-demographic $d$: $\text{P}_i(d)$ from Equation \ref{eq:weight}.

\textbf{Post-stratification.}
Post-stratification reweights each user according to the joint distribution of a set of socio-demographics~\cite{holt1979post,little1993post,henry2012methods}. 
In practice, this joint distribution is rarely known or available to researchers beyond two or three variable combinations.
The two methods below address this situation and use only the marginal distributions for each socio-demographic.

\textbf{Naive Post-stratification.}
Since joint distributions are not always available for many variables of interest, one can estimate the joint distribution from given marginals. One approach is to assume all marginal distributions are independent~\cite{leemann2017extending}. This method multiplies the proportion of people in each marginal bin to estimate the proportion of people in each of the joint distribution's bins, mirroring the assumption of Naive Bayes ($p(a, b) = p(a)p(b)$).

\textbf{Raking.}
Raking is an iterative method which operates on the marginal distributions, adjusting each sample marginal to match the population distributions~\cite{deville1993generalized}. 
For example, raking over age and gender would first adjust age sample marginals to match age population marginals, and then adjust gender sample marginals to match gender population marginals. This process is repeated until the marginal distributions of the sample variables match the population marginal distributions within some small margin of error. 
The adjusted sample marginals are then substituted into the numerator of Equation \ref{eq:weight}.

\subsection{Applying Weight Factors and Predictive Modeling}

We apply our correction weights to individual level (Twitter users) linguistic features, specifically the top 25,000 most frequent unigrams across our entire sample, noting that this procedure will work for any individual level data. We concatenate all tweets from each user in our data set and tokenize using a tokenzier built for social media data~\cite{dlatk2017emnlp}. We then encode each unigram the relative frequency of use for each given user. 
Using Equation \ref{eq:pop average}, each linguistic feature $x_{j}$ is aggregated from user $j$ to county $i$:
\begin{equation}
    \hat{\mu}_{X_i} = \frac{1}{N_i}\sum_{j\in U_i} \hat{\psi}_i(d)r_{j}(x_j).
    \label{eq:feature average}
\end{equation}
Here $U_i$ is the set of users in county $i$, $N_i$ is the total number of Twitter users in county $i$, $\hat{\psi}_i(d)$ is the correction weight of the demographic set $d$, and $r_j(x_j)$ is the relative frequency of the unigram $x_j$ for user $j$. When aggregating from user to county with no bias correction we set $\hat{\psi}_i(d)=1, \forall j\in U_i$ and $\forall i$.
The end results is a set of 25,000 county-level average unigrams.

\subsection{Predictive Modeling}
\label{section:modeling details}

Since our methods focus is on selection bias, we integrate our correction approach into an established approach for estimating county-level health statistics from language   ~\citep{eichstaedt2015psychological,giorgi2018remarkable,jaidka2020estimating}.

\paragraph{Features} 
The county-level average unigrams are then used to derive a set of topic loadings for each county. 
We use a set of 2,000 topics derived from Latent Dirichlet Allocation~\cite{blei2003latent}.
The topics were built over the myPersonality Facebook data set, which consists of approximately 15 million; see ~\citet{schwartz2013personality} for more details on the topic modeling process.
These topics have been used across a number of studies predicting county-level health and well-being~\cite{eichstaedt2015psychological,giorgi2018remarkable,curtis2018can}.

\paragraph{Modeling} For each of our four county-level prediction tasks, we predict the outcome (i.e., heart disease mortality, suicide mortality, life satisfaction, and percentage in poor or fair health) using the 2,000 topic features described above in a 10-fold cross validation setup. 
Thus, the final data size is dependent on the county outcome (i.e., dependent variable), with 2,000 LDA topics (i.e., independent variables) constant across each task: heart disease mortality $N=2,038$; suicide mortality $N=1,672$; life satisfaction $N=1,951$; and percentage in poor or fair health $N=1,931$.
The topic features are then fed through a three step feature selection pipeline.
First, we remove all low variance features.
Next, we remove features which are not correlated with our county-level outcomes at a family-wise error rate $\alpha$ of 60.
At this point, all features are standardized: mean centered and normalized by the standard deviation.
Finally, stochastic principal component analysis is applied to the topic features in order to reduce the size of the feature set to approximately 10\% its original size.
For each of the 10 folds, we train an $\ell_2$ penalized ridge regression~\cite{hoerl1970ridge} on 9 of the folds and apply the model to the held out 10th fold.
All models use a regularization term $\lambda$ of 10,000.
Similarly to the topic feature set, this same pipeline has been successfully used across a number of county-level health studies~\cite{eichstaedt2015psychological,giorgi2018remarkable,jaidka2020estimating,abebegiorgi2020quantifying}


\subsection{Results for Out-of-the-box Methods}
\label{section:results_selection_bias}

In this section we evaluate how well existing post-stratification techniques improve prediction accuracy by correcting for selection bias. 
We focus on the average cross-validation accuracy across the four health outcomes introduced previously: heart disease mortality,  suicide mortality, life satisfaction, and percent in poor or fair health. 
The assumption is that if a mitigation technique is useful it should improve predictive performance, while unnecessary or erroneous techniques will have no or negative effect. 


\paragraph*{Predictive Performance}

\begin{table}[t]\centering\fontsize{9}{9}\selectfont
\begin{tabular}{lcccc|c}\toprule
 & \multirow{2}{*}{\begin{tabular}[c]{@{}c@{}}Heart \\ Disease\end{tabular}} & \multirow{2}{*}{\begin{tabular}[c]{@{}c@{}}Suicide\end{tabular}} & \multirow{2}{*}{\begin{tabular}[c]{@{}c@{}}Life \\ Sat.\end{tabular}} & \multirow{2}{*}{\begin{tabular}[c]{@{}c@{}}Poor/Fair \\ Health\end{tabular}}  & \multirow{2}{*}{\begin{tabular}[c]{@{}c@{}}Avg.\end{tabular}}  \\ 
  &  &  &  &  &   \\ \hline
Baseline & .751 & .614 & .445 & .748 & .640  \\ \hline
Age & .622$^-$ & .492$^-$ & .239$^-$ & .595$^-$ & .487$^-$  \\
Gender & .755$^+$ & .619$^+$ & .437$^-$ & .744 & .641 \\
Income & .703$^-$ & .530$^-$ & .455 & .716$^-$ & .599$^-$ \\ 
Education & .762$^+$ & .617 & .457$^+$ & .754$^+$ &  .648$^+$ \\
\bottomrule
\end{tabular}
\caption{Standard methods using single correction factors. Reported Pearson r., $^+$ and $^-$ indicate a significant increase and decrease in performance, respectively ($p<0.05$). Half of the ``out-of-the-box" correction factors reduce accuracy.}
\label{table:initial single factor results}
\end{table}

As shown in Table \ref{table:initial single factor results}, we found a decrease in performance (Pearson $r$) when attempting to correct for both age and income biases. Gender correction has mixed results, as we see an increase for heart disease and suicide, a decrease for life satisfaction and no change for poor/fair health. Education gives a significant boost for three out of four tasks, with no change for suicide. For each correction factor we perform a pair t-test on the model's residual as compared to the baseline model residual. We report both positive ($^+$) and negative ($^-$) statistical differences at $p<0.05$.

Similar patterns hold when averaging across all four tasks. We note that at this point in the paper we will report the average Pearson r across all four tasks (heart disease, suicide, life satisfaction, and poor/fair health). Additionally, tests for significance are done by combining the dependent p-values across the four county-level tasks using the methods developed by Kost and McDermott~\cite{kost2002combining}.
We also note that age and income have 10 and 11 possible bins, respectively, whereas both gender and education are binary variables.
An increased number of bins can lead to more extreme weights if any bins are densely or sparsely populated, thus increasing the noise in our model. 
This suggests some issues perhaps arising from having many bins (e.g. sparse or unstable estimates of people per bin; we will address this in the next two sections).


Average predictive performance for combinations of correction factors is given in Table \ref{table:initial combined factor results}. Again, across the board we see no increase in predictive performance when comparing to baseline. Additionally, we see no increase in predictive performance when comparing to a single factor correction in Table \ref{table:initial single factor results}. 

\begin{table}[tb!]\centering
\begin{tabular}{lcc}\toprule
 \textbf{\emph{Baseline} $=.640$}  & \begin{tabular}[c]{@{}c@{}}Naive \\ Post-Stratification\end{tabular} & Raking \\ \hline
Age + Gender & .514$^-$ & .473$^-$ \\
Income + Education & .600$^-$ & .591$^-$ \\
Age + Gen. + Inc. + Edu. & .627$^-$ & .541$^-$ \\
\bottomrule
\end{tabular}
\caption{Standard methods using multiple correction factors, average predictive accuracy (Pearson r) across four tasks, $^-$ significant decrease in performance at $p<0.05$. All six ``out-of-the-box" combinations reduce predictive accuracy.}
\label{table:initial combined factor results}
\end{table}

\section{Study 2: \name for Improved Correction}
\label{section:methods_challenges}

Standard selection bias mitigation techniques not only provided no benefit, but, on average, tended to hurt performance within the context of predicting county health from social media language. 
We hypothesize this is due to two challenges.
First, socio-demographic estimation from language introduces systematic effects to the distributions (e.g. from shrinkage -- bias toward the mean).  
Second, data sparsity is an issue when dealing multi-dimensional socio-demographics, since some counties contain as few as 100 individuals. 

\subsection{Methods}

\begin{algorithm}[b!]
\DontPrintSemicolon
    \SetKwInput{KwInput}{Input}                
    \SetKwInput{KwOutput}{Output}              
    \KwInput{$\lbrace d^{(s)}\rbrace$ - demographic estimates from users 
    $expectedBinPercs$ - Expected percentages} 
    \KwOutput{$\lbrace d^{(t)}\rbrace$ - redistributed demographic estimates from users}

  \SetKwFunction{FAB}{EstRedist}
  \SetKwProg{Fn}{Def}{:}{}
  \Fn{\FAB{$\lbrace d^{(s)}\rbrace$, $expectedBinPercs$}}{
        \For{h \text{in} length(expectedBinPercs)}    
        { 
            p = percentBetween($\min^{(t)}_h$, $\max^{(t)}_h$)\;
            \If{l == 0}{
                $\min^{(s)}_h$ = $\min^{(t)}_h$\;
            } 
            \Else{
                $\min^{(s)}_h$ = $\max^{(s)}_{h-1}$\;
            }
            $\max^{(s)}_h$ = $\min^{(s)}_h$ + 1\;
            
            \While{percentBetween($\min^{(s)}_h$, $\max^{(s)}_h) < p$}{
                $\max^{(s)}_h$ += 1\;
            }

            \For{$\lbrace d^{(s)}\rbrace$}{
                \If{$\min^{(s)}_h$ $\leq$ $d^{(s)}$ $<$ $\max^{(s)}_h$}{
                    $d^{(t)}$ = EquationSix($d^{(s)}$)\;
                }
            }
        }
        \KwRet $\lbrace d_t\rbrace$
    }
    \caption{\textit{Estimator Redistribution}}
    \label{alg:estimator redistribution}
\end{algorithm}

\paragraph*{Challenge 1: Estimator Shrinking}
The first challenge originates in Step 1 of our pipeline: estimating socio-demographics from text.
The estimators used to create the linguistic socio-demographic scores are regularized which shrinks the estimated distribution towards the mean of the training data. 
To compensate for this, each users' estimated socio-demographics are redistributed such that our source distribution matches that of a target distribution, in our case, that of the national distribution of Twitter users ($expectedBinPercs$, as reported by the PEW Reseach Center). 
\textbf{\textit{Estimator redistribution}} shifts each acount's linguistic socio-demographic estimates such that the population percentage in each source bin matches those of the target bins. 
Specifically, for a given socio-demographics bin $h$, the bin boundaries in the source data ($\text{min}^{(s)}_h$ and $\text{max}^{(s)}_h$) were determined such that they match proportions in target population distribution bins ($\text{min}^{(t)}_h$ and $\text{max}^{(t)}_h$). See Algorithm \ref{alg:estimator redistribution} for details and the Supplemental Materials\footref{supp_and_data} for a step-by-step example of this method.

\begin{figure}[t]
\centering
\includegraphics[width=.35\textwidth]{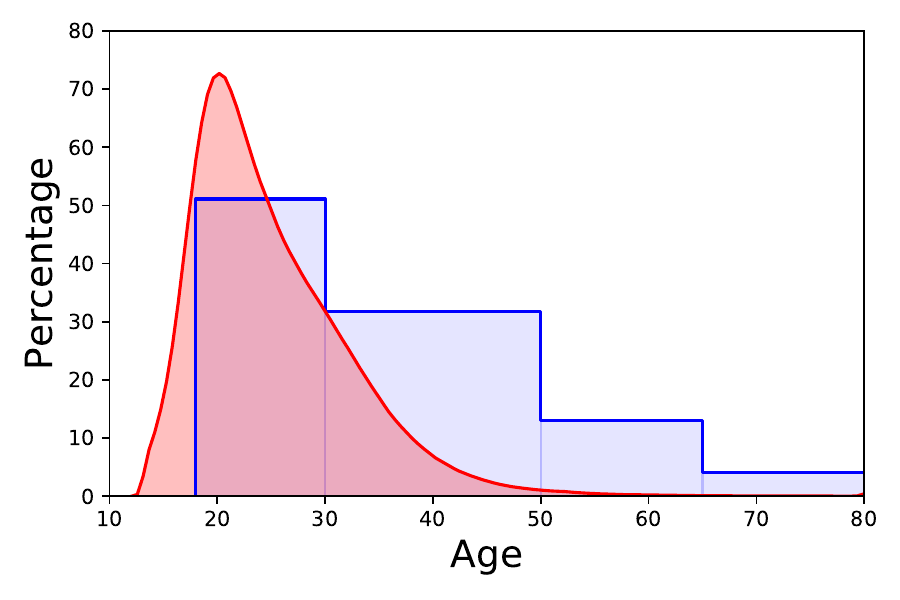}
\caption{Probability density of the age distributions of our \emph{Twitter sample} (red) versus the expected \emph{Twitter population} distribution according to PEW (blue). Due to regularization which shrinks estimates toward the mean, our age distribution skews significantly younger than PEW.
}
\label{fig:distributions}
\end{figure}

A given user's estimated socio-demographic $d^{(s)}$ (where $s$ is the source distribution) is redistributed using the following equation:
\begin{equation}
    \frac{d^{(s)} - \text{min}^{(s)}_h}{\text{max}^{(s)}_h - \text{min}^{(s)}_h} = \frac{d^{(t)} - \text{min}^{(t)}_h}{\text{max}^{(t)}_h - \text{min}^{(t)}_h},
\nonumber
\end{equation}
The redistributed estimation value is obtained by solving for $d^{(t)}$, the socio-demographic in the target distribution $t$:
\begin{equation}
   d^{(t)} = \big(d^{(s)} - \text{min}^{(s)}_h\big) \frac{\text{max}^{(t)}_h - \text{min}^{(t)}_h}{\text{max}^{(s)}_h - \text{min}^{(s)}_h} + \text{min}^{(t)}_h.
\label{eq:redistribution}
\end{equation} Figure \ref{fig:distributions} shows the age distributions of our national Twitter sample and PEW's reported national percentages. 

We expect estimator redistribution to help when there is a large number of socio-demographic bins and when there exists large differences between the sample and target distributions, regardless of the number of bins. The redistribution process will move users from densely populated bins into sparser bins, yielding more stable correction factors --- users in extreme bins (either dense or sparse) are severely under or over-weighted.

\paragraph*{Challenge 2: Sparse Data Bins}

The second challenge originates in Step 2 of our pipeline: creating weight factors. 
As we (1) increase the number of bins of our socio-demographic variables and (2) increase the number of socio-demographic variables we wish to correct for, the probability that any one of our sample users falls into a given bin also shrinks. 
As seen in Equation \ref{eq:weight}, as the percentage of users in our sample shrinks, weights will increase. 
The raking process described above also suffers from the fact that convergence is not guaranteed if empty bins are present~\cite{battaglia2009practical}. 
Therefore we focus on ways of estimating $\text{Q}_i(d)$ in Equation \ref{eq:weight} such that we mitigate this data sparsity problem. 

\paragraph{\textbf{Adaptive Binning}}
Our first method to account for sparse data sets a minimum threshold on the number of observations within each bin (or partition subset) for a given socio-demographic variable. 
Adjacent bins are iteratively combined until all bins meet our threshold or we have a single bin. 
Since both gender and education start with two bins, if either bin fails to meet the threshold then we end up with a single bin and therefore no correction.
Thus, we do not expect either variable to significantly increase or decrease predictive performance from baseline.
We also note that adaptive binning occurs per socio-demographic (e.g., when correcting for both age and income, we bin age and income separately).  
While a minimum bin threshold has previously been used (\citet{battaglia2009practical} who suggest a minimum bin percentage of 5\%), we know of no systematic study of the effect of minimum bin sizes. Additionally, our threshold is set on the number of observations as opposed to a percentage, since percentages will be noisy for sparsely populated counties.  
The Adaptive Binning algorithm is shown in Algorithm \ref{alg:adaptive binning}; see Supplemental Materials\footref{supp_and_data} for a step-by-step example of this method. 

\begin{algorithm}[b!]
\DontPrintSemicolon
    \SetKwInput{KwInput}{Input}                
    \SetKwInput{KwOutput}{Output}              
    \KwInput{$binCounts$ - list of bin counts $minBinNumber$ - min integer bin threshold $binRanges$ - list of bin ranges}
    \KwOutput{$binCounts$ - list of combined bin counts $binRanges$ - list of combined bin ranges}
  \SetKwFunction{FAB}{AdaptBin}
  \SetKwProg{Fn}{Def}{:}{}
  \Fn{\FAB{$binCounts$, $minBinNumber$}}{
        \While{min(binCounts) $<$ minBinNumber}    
        { 
            m = min{(binCounts)}\;
            \If{m $\geq$ minBinNumber}{\text{break}\;}
            i = binCounts.index\_of(m)\; 
            combineAdjacent(binCounts[i], min(binCounts[i-1],binCounts[i+1]))\;
            combineAdjacent($binRanges$[i], min($binRanges$[i-1],$binRanges$[i+1]))\;
        }
        \KwRet $binCounts$, $binRanges$
    }
    \caption{Adaptive Binning}
    \label{alg:adaptive binning}
\end{algorithm}

\paragraph{\textbf{Informed Smoothing}}
The second method we develop to account for data sparsity uses a smoothing technique that pads each weight with a fraction of users from a known distribution.  
More formally, we state the source probability in terms of the smoothing constant $k$ as
\begin{equation}
    \hat{\text{Q}}_i^{(k)}(d|d_m\in D^{(h)}_{d_m}, \forall m) = \frac{N_s + k\hat{\text{P}}_i(d)}{N_i + k} .
    \label{eq:informed smoothing}
\end{equation}
Here $N_s$ is the number of sample users with socio-demographic $d$, $N_i$ is the number of Twitter users in county $i$ and $h$ is summed over the socio-demographic partition. 
Note that as $k \to\infty$ as have $\hat{\text{Q}}_i^{(k)}(d)\to \hat{\text{P}}_i(d)$ and therefore all correction weight factors equal 1.


Unlike adaptive binning, informed smoothing does not depend on the total number of bins. Thus, we expect it to have an effect on gender and education correction. 
This approach is inspired by similar approaches to modeling ngram probabilities in language modelling~\cite{kneser1995improved}.

  

 



\begin{figure}[tb]
\centering\includegraphics[width=.9\columnwidth]{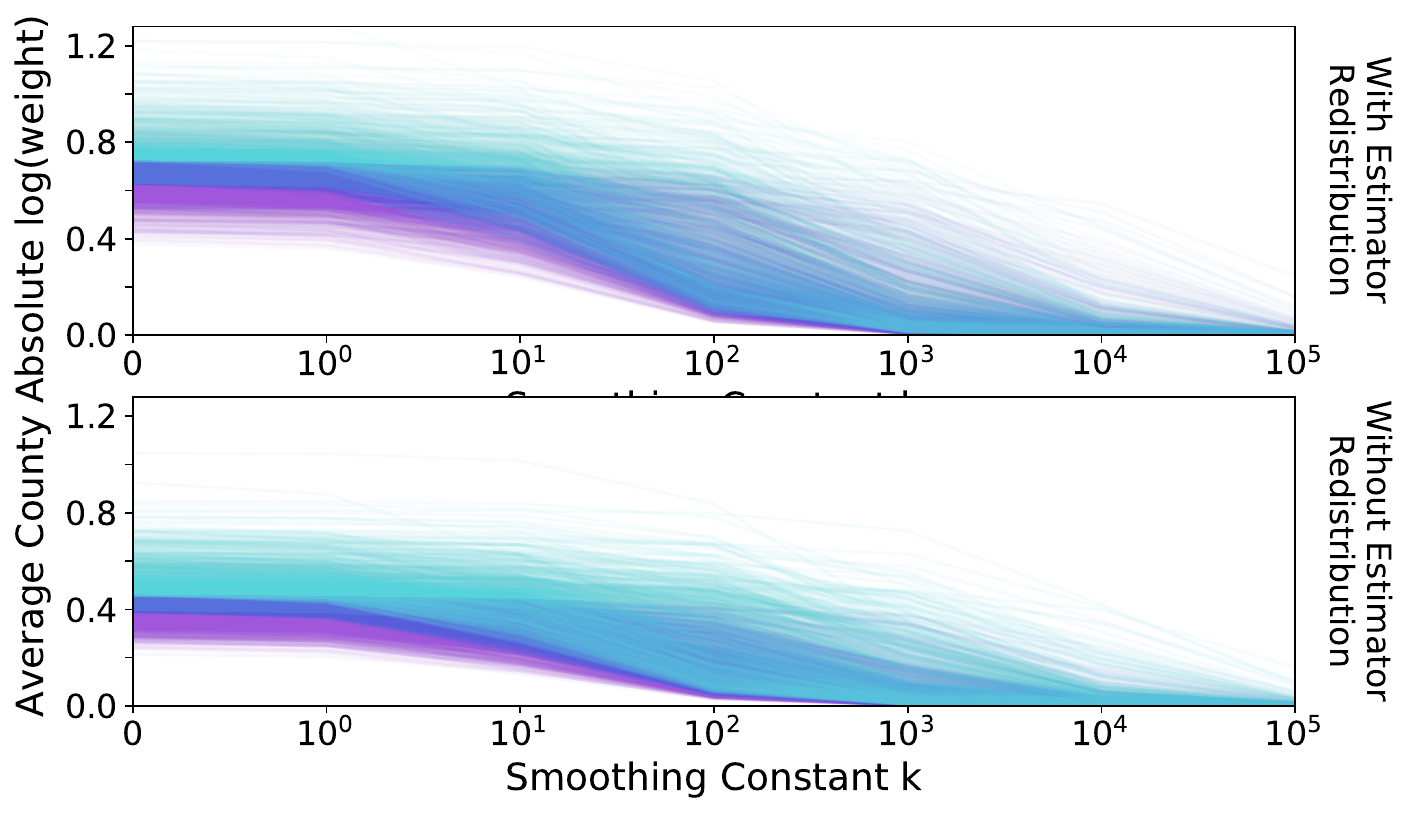}
\caption{Average, absolute log of the county income weights at different smoothing levels; colored by terciles with (top) and without (bottom) estimator redistribution. }
\label{fig:renorm}
\end{figure}

\subsection{Results: Estimation and Sparsity Challenges}
\label{section:results_challenges}

To be sure our methods work as expected, we first observe their effects on the correction weights. Figure \ref{fig:renorm} shows results for both estimator redistribution and informed smoothing for income alone. Each figure shows the county average, absolute log of the users' correction factors. First, ignoring the effects of smoothing (i.e., focusing on $k=0$), we see that estimator redistribution shrinks the variance in the correction factors. This is to be expected since estimator redistribution spreads out the distribution --- the sample  distribution (red) in Figure \ref{fig:distributions} is spread to match the true distribution (blue). This causes our bins to (1) be less sparse near the tail of the distribution (thus, shrinking large correction factors towards the mean); and (2) less dense near the peak of the distribution (similarly, increasing small correction factors towards the mean). 

Informed smoothing also has a similar shrinking effect at large $k$, with average log weights converging to zero. 
This is expected since Equation \ref{eq:informed smoothing} says that, as $k$ increases, the estimated sample distribution $\hat{\text{Q}}_i^{(k)}(d)$ matches the estimated population distribution $\hat{\text{P}}_i(d)$. Thus, $\hat{\psi}_i(d)\to 1$ and the log approaches 0.
Finally, we see that the variance in weights does not monotonically decrease, with maximum variance at $k=100$.
At $k=100$ we see the terciles calculated at $k=0$ spreading out. 
Since we are correcting weights on a county-by-county basis, with each county having it's own selection bias, we would hope that the average county weights show variance.  
At $k=0$ we see this is not the case and all counties have similar average weights, implying that all counties are experiencing the same selection bias (i.e., a similar ratio of population to sample; Equation \ref{eq:weight}).

Table \ref{table:redistribution selection bias results} evaluates the benefit of applying estimator redistribution. Comparing to Tables \ref{table:initial single factor results} and \ref{table:initial combined factor results}, we see a marked improvement above post-stratification without estimator redistribution in almost all situations. It does not put us above baseline ($r = .640$) but it is moving in the right direction, so we use estimator redistribution in all remaining experiments. 

\begin{table}[tb]\centering\small
\begin{tabular}{lccc}\toprule
  \textbf{\emph{Baseline} $=.640$} & \begin{tabular}[c]{@{}c@{}}Post- \\ Stratification\end{tabular} & \begin{tabular}[c]{@{}c@{}}Naive Post-\\ Stratification\end{tabular} & Raking \\ \hline
Age & .587$^+$ & - & - \\
Gender & .639 & - & - \\
Income & .625$^+$ & - & - \\
Education & .648 & - & - \\
Age + Gender & - & .582$^+$ & .584$^+$ \\
Inc. + Edu. & - & .629$^+$ & .626$^+$ \\
All & - & .638$^+$ & .588$^+$ \\ 
\bottomrule
\end{tabular}
\caption{Average predictive accuracies (Pearson r) across four tasks when using \textit{estimator redistribution}. $^+$ and $^-$ indicate a significant increase or decrease, respectively, as compared to same correction variable / method pair in Tables \ref{table:initial single factor results} and \ref{table:initial combined factor results}. All correction factors, except \emph{gender} and \emph{education}, show significant increases over the ``out-of-the-box" methods. ``All" includes age, gender, income, and education.}
\label{table:redistribution selection bias results}
\end{table}

The predictive accuracies for the adaptive binning experiments are shown in Table \ref{table:adaptive binning}. 
This marked our first improvement over the baseline average Pearson r of $.640$. 
In most cases, we see a decrease in performance when setting the minimum bin threshold to 1 when compared to no binning. 
We also see naive combining outperforming raking when $k$ is low, though $k\geq50$ reverses this and raking outperforms naive. 
Increasing the minimum bin threshold gradually improved results, with peaks around 100 where all approaches did better than no adaptive binning.
As expected, most factors approach baseline when the bin threshold is 1,000.

\begin{table}[b!]\centering\small
\begin{tabular}{lccccc} \toprule
\multirow{2}{*}{\textbf{\emph{Baseline}} = .640} & \multicolumn{5}{c}{Minimum Count Threshold} \\ \cline{2-6}
 & 1 & 10 & 50 & 100  & 1000 \\ \hline

Age & .583 & .605$^+$ & .624$^+$ & .636$^+$  & .634$^+$ \\
Gender & .639 & .639 & .639 & .640 & .640 \\
Income & .612$^-$ & .666$^\star$ & .674$^\star$ & .663$^\star$ & .642$^+$ \\
Education & .648$^\star$ & .648$^\star$ & .648$^\star$ & .647$^\star$ & .642 \\
Age + Gender & & & &  &    \\
\hspace{3mm}Naive & .580 & .598$^+$ & .622$^+$ & .633$^+$ & .633$^+$ \\
\hspace{3mm}Raking & .580 & .603$^+$ & .623$^+$ & .635$^+$ & .634$^+$ \\
Inc. + Edu. & & & & &   \\
\hspace{3mm}Naive & .612$^-$ & .659$^\star$ & .673$^\star$ & .662$^\star$ & .643 \\
\hspace{3mm}Raking & .611$^-$ & .662$^\star$ & .674$^\star$ & .664$^\star$ & .643 \\
All & & & & &  \\
\hspace{3mm}Naive & .634 & .633 & .620$^-$ & .634 & .645$^+$ \\
\hspace{3mm}Raking & .579$^-$ & .610$^+$ & .634$^+$ & .649$^\star$ & .647$^\star$ \\ 
\bottomrule
\end{tabular}
\caption{Average predictive accuracies (Pearson r) across four tasks when using \textit{adaptive binning}. $^+$ and $^-$ indicate a significant increase or decrease, respectively, as compared to the same correction variable / method pair in Table \ref{table:redistribution selection bias results}, $^\star$ increase over baseline. ``All" includes age, gender, income, and education. This method shows mitigating selection bias can improve predictive accuracy when adjusting for error in demographic scores by using adaptive binning.}
\label{table:adaptive binning}
\end{table}

\begin{figure*}[ht]
\centering
\begin{tabular}{ccc}
  \includegraphics[width=.55\columnwidth]{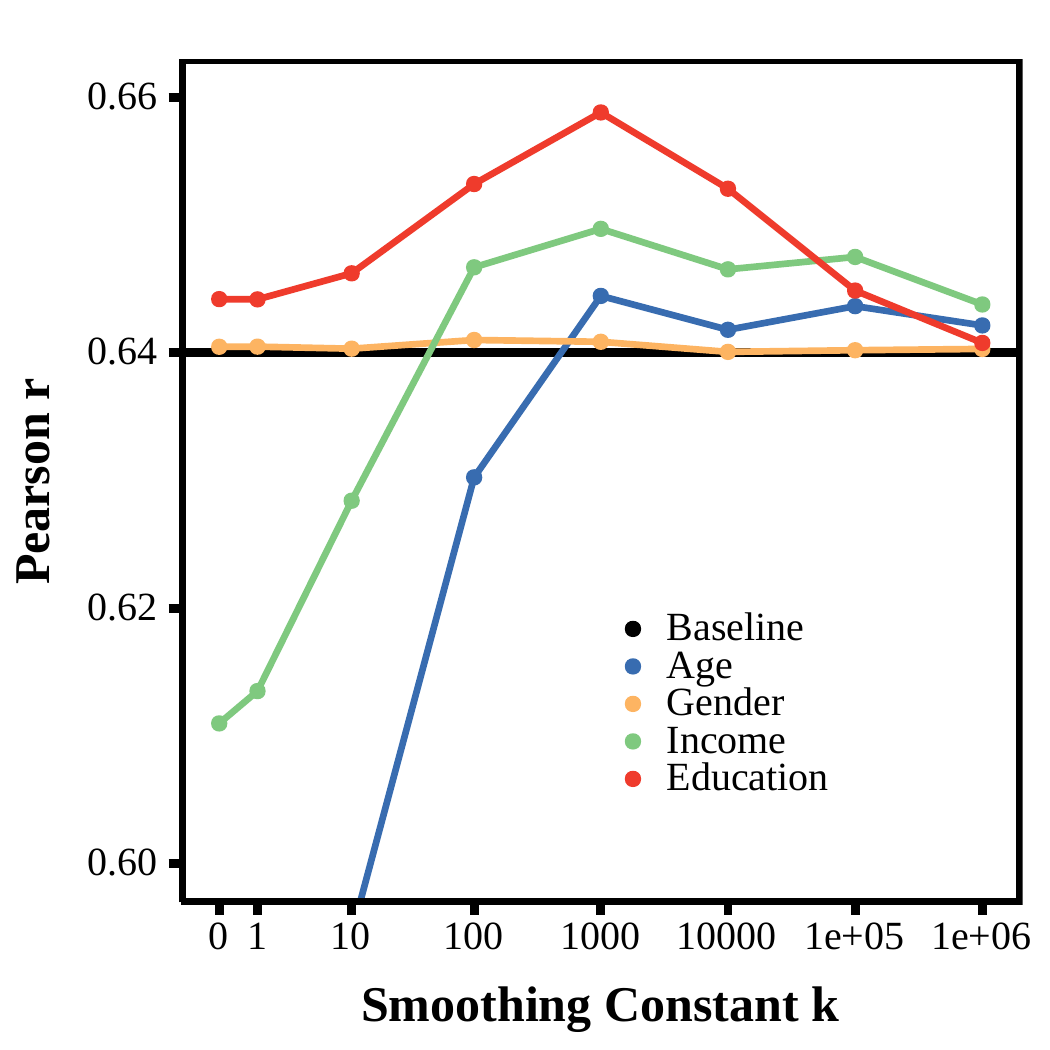}\label{fig:smoothing and binning a} & \includegraphics[width=.55\columnwidth]{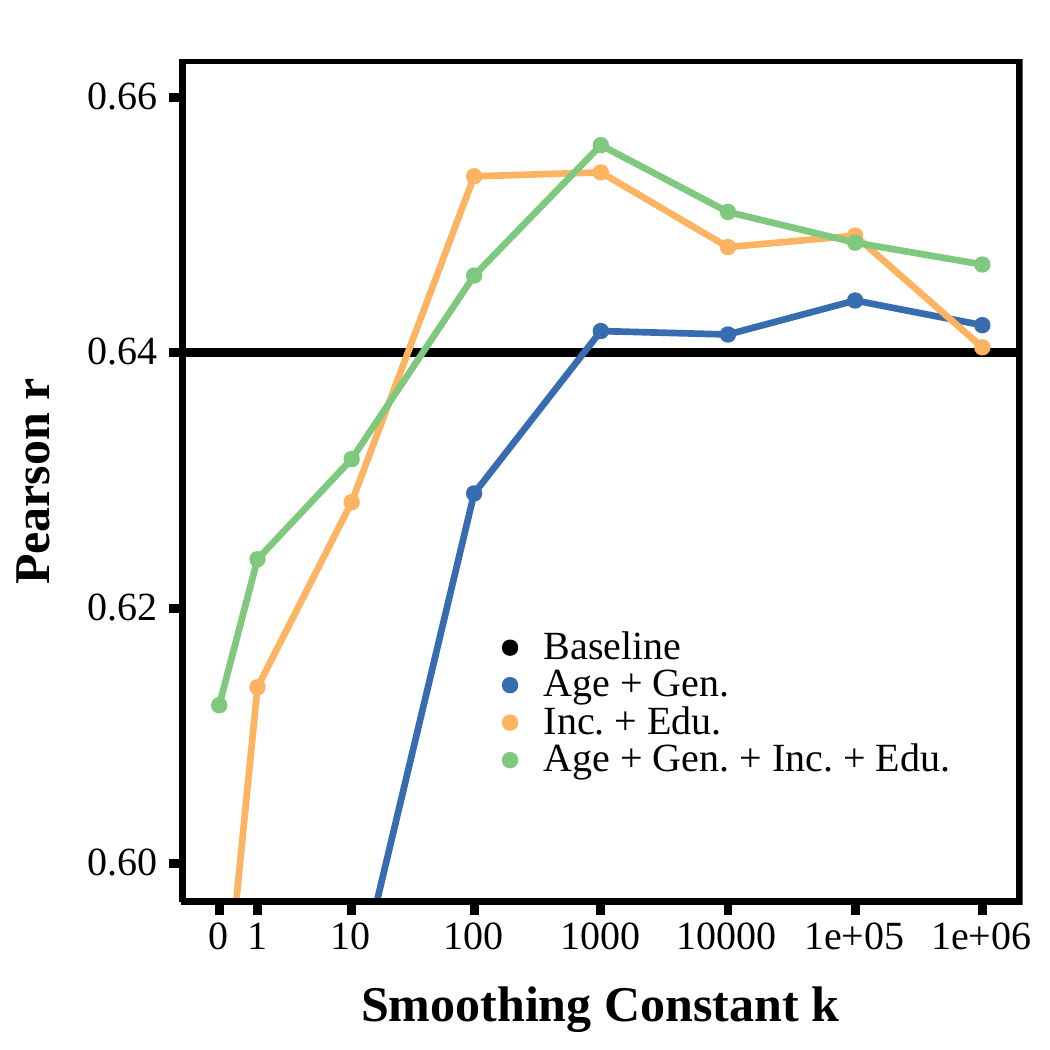}\label{fig:smoothing and binning b} & \includegraphics[width=.55\columnwidth]{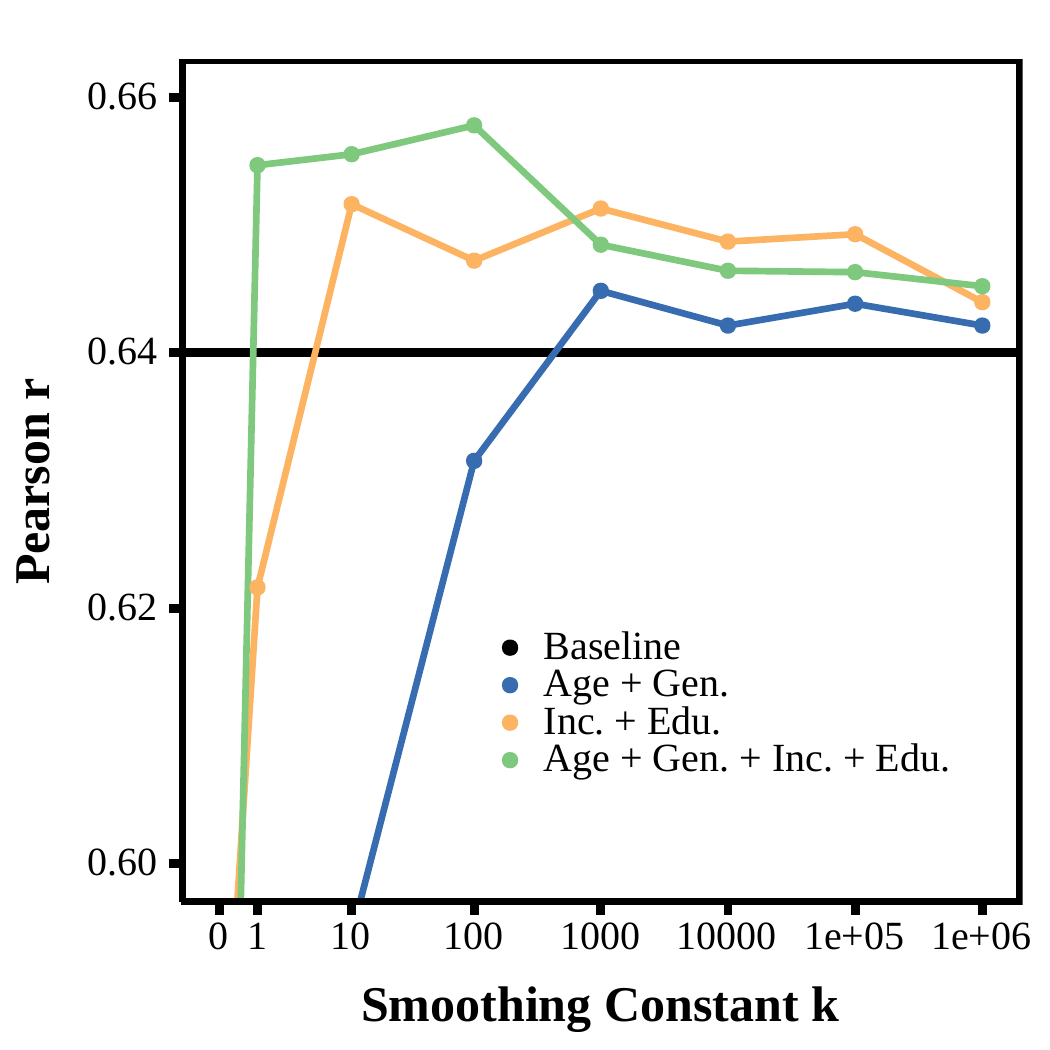}  \\
(a) Single Factors & (b) Naive Factors & (c) Raking Factors  \\[6pt] 
\end{tabular}
\caption{Prediction accuracies when using \textit{informed smoothing}, averaged over all four tasks: Graphs zoomed in to highlight difference near baseline; smoothing constant $k=0$ is equivalent to no smoothing (see Table \ref{table:redistribution selection bias results}). Results show selection bias mitigating, when adjusted with informed smoothing with $k > 10$, can increase accuracy over baseline. }
\label{fig:smoothing}
\end{figure*}

Figure \ref{fig:smoothing} shows the predictive accuracies of the informed smoothing method. Figure \ref{fig:smoothing}(a) shows informed smoothing with single correction factors. 
For single factors alone, we see a slight increase using age and larger increases for income and education.
Consistent with our previous results, we see no improvements for gender correction.
All results converge to no correction with large enough $k$ since the Informed Smoothing has the effect of backing off to assuming the county is fully representative (i.e. no correction).\footnote{Uninformed smoothing, such as Lapacian smoothing, would push counties toward a non-representative (uniform) distribution, negating the point of selection bias correction. See Supplement\footref{supp_and_data} for an ``add one" smoothing.} 

Figures \ref{fig:smoothing}(b) and \ref{fig:smoothing}(c) show Informed Smoothing with naive and raking factors, respectively. 
We see that the combination of age and gender does not drastically improve over baseline.
We also see raking helping more for the age-gender-income-education correction factor (for both Informed Smoothing and Adaptive Binning), suggesting that raking might work better than naive post-stratification as the number of correction factors increases.

\begin{figure}[b!]\centering
\begin{tabular}{c}
  \includegraphics[width=.8\columnwidth]{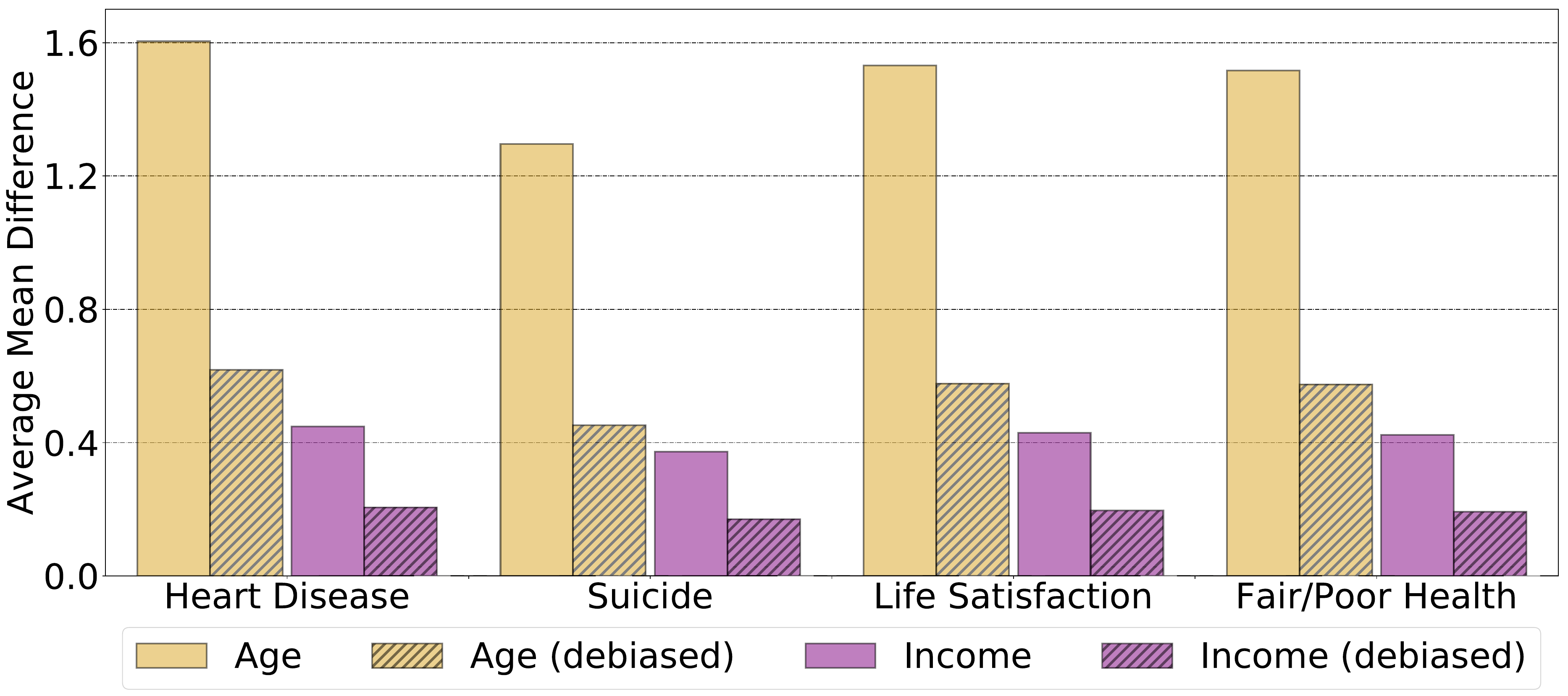}\label{fig:quantifying bias a} \\
(a) Continuous Correction Factors \\ \includegraphics[width=.8\columnwidth]{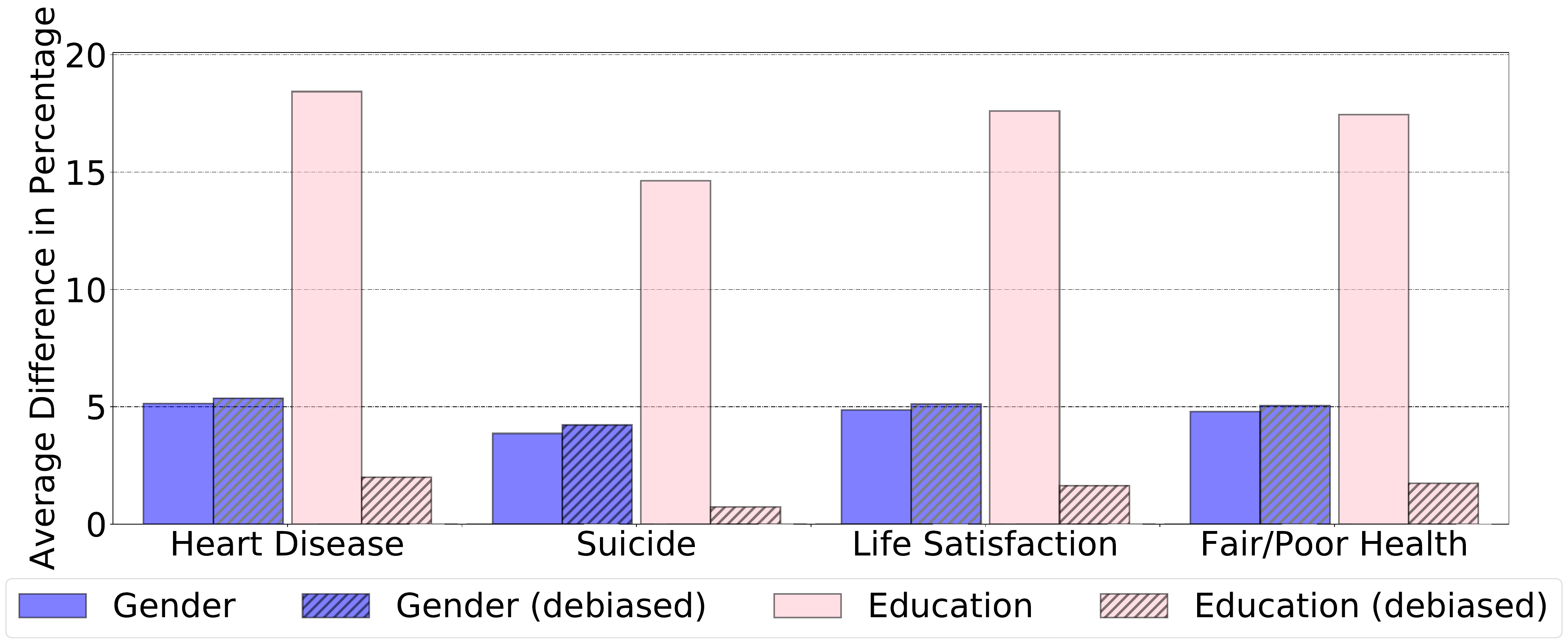}\label{fig:quantifying bias b}  \\ (b) Categorical Correction Factors  \\[6pt] 
\end{tabular}
\caption{Effect of mitigation on average bias. Bars indicate bias for both (a) continuous attributes (age and income) quantified as difference in standardized means, or (b) dichotomous attributes (gender and education) quantified as percentage difference between census populations and our measurements. Darker bars indicate the debiased version, from our final suggested approach using estimator redistribution, adaptive binning, and informed smoothing -- settings from the last line of Table \ref{table:best model}). Bias is reduced for all factors except gender (which only had a small baseline bias).  }
\label{fig:quantifying bias}
\end{figure}

\begin{table*}[h]\centering
\begin{tabular}{lccccccccc}\toprule
 & \multicolumn{3}{c}{Baseline} & \multicolumn{6}{c}{Optimal Model based on Backwards Elimination} \\ \cmidrule(lr){2-4}  \cmidrule(lr){5-10} 
 & \multirow{2}{*}{\begin{tabular}[c]{@{}c@{}}Pearson \\ r\end{tabular}} & \multirow{2}{*}{\begin{tabular}[c]{@{}c@{}}R$^2$ \end{tabular}} & \multirow{2}{*}{\begin{tabular}[c]{@{}c@{}}\textit{RMSE} \end{tabular}} & \multirow{2}{*}{\begin{tabular}[c]{@{}c@{}}Strat. \\ Vars.\end{tabular}}  & \multirow{2}{*}{\begin{tabular}[c]{@{}c@{}}Adaptive \\ Binning\end{tabular}} & \multirow{2}{*}{\begin{tabular}[c]{@{}c@{}}Smoothing \\ k\end{tabular}} & \multirow{2}{*}{\begin{tabular}[c]{@{}c@{}}Pearson \\ r\end{tabular}} & \multirow{2}{*}{\begin{tabular}[c]{@{}c@{}}R$^2$ \end{tabular}} & \multirow{2}{*}{\begin{tabular}[c]{@{}c@{}}\textit{RMSE} \end{tabular}} \\ 
    & &  &  &  &  &  &    \\ \hline
Heart Disease & .753 & .563 & \textit{30.14} & inc. + edu. & 50 & 10 & .769$^*$  & .588 & \textit{29.26} \\
Suicide & .614 & .371 & \textit{3.67} & inc. & 50 & 10 & .626$^*$  & .388 & \textit{3.62} \\
Life Satisfaction & .445 & .187 & \textit{0.024} & inc. & 50 & 10 & .542$^*$  & .286 & \textit{0.022} \\
Poor/Fair Health  & .746 & .552 & \textit{3.82} & inc. + edu. & 50 & 10 & .778$^*$  & .603 & \textit{3.60} \\ 
\hline
\multirow{2}{*}{\begin{tabular}[c]{@{}c@{}}Average\end{tabular}} & \multirow{2}{*}{\begin{tabular}[c]{@{}c@{}}.640\end{tabular}} & \multirow{2}{*}{\begin{tabular}[c]{@{}c@{}}.418\end{tabular}} & \multirow{2}{*}{\begin{tabular}[c]{@{}c@{}}\textit{9.41}\end{tabular}}  & inc. & 50 & 10 & .677$^*$  & .464 & \textit{9.18}\\ 
  & &  &  & inc + edu & 50 & 10 & .676$^*$ & .463 & \textit{9.13} \\ 
\bottomrule
\end{tabular}
\caption{Predictive performance of baseline vs. our recommended system. Across each of our four individual county-level prediction tasks we present a recommended system (i.e., the combination of settings which resulted in the highest prediction accuracy with the smallest number of correction factors). $^*$ indicates significant reduction in error over baseline; $p<.05$.}
\label{table:best model}
\end{table*}

\begin{figure*}[ht!]
\begin{subfigure}{.5\textwidth}
  \centering
  \includegraphics[width=.6\linewidth]{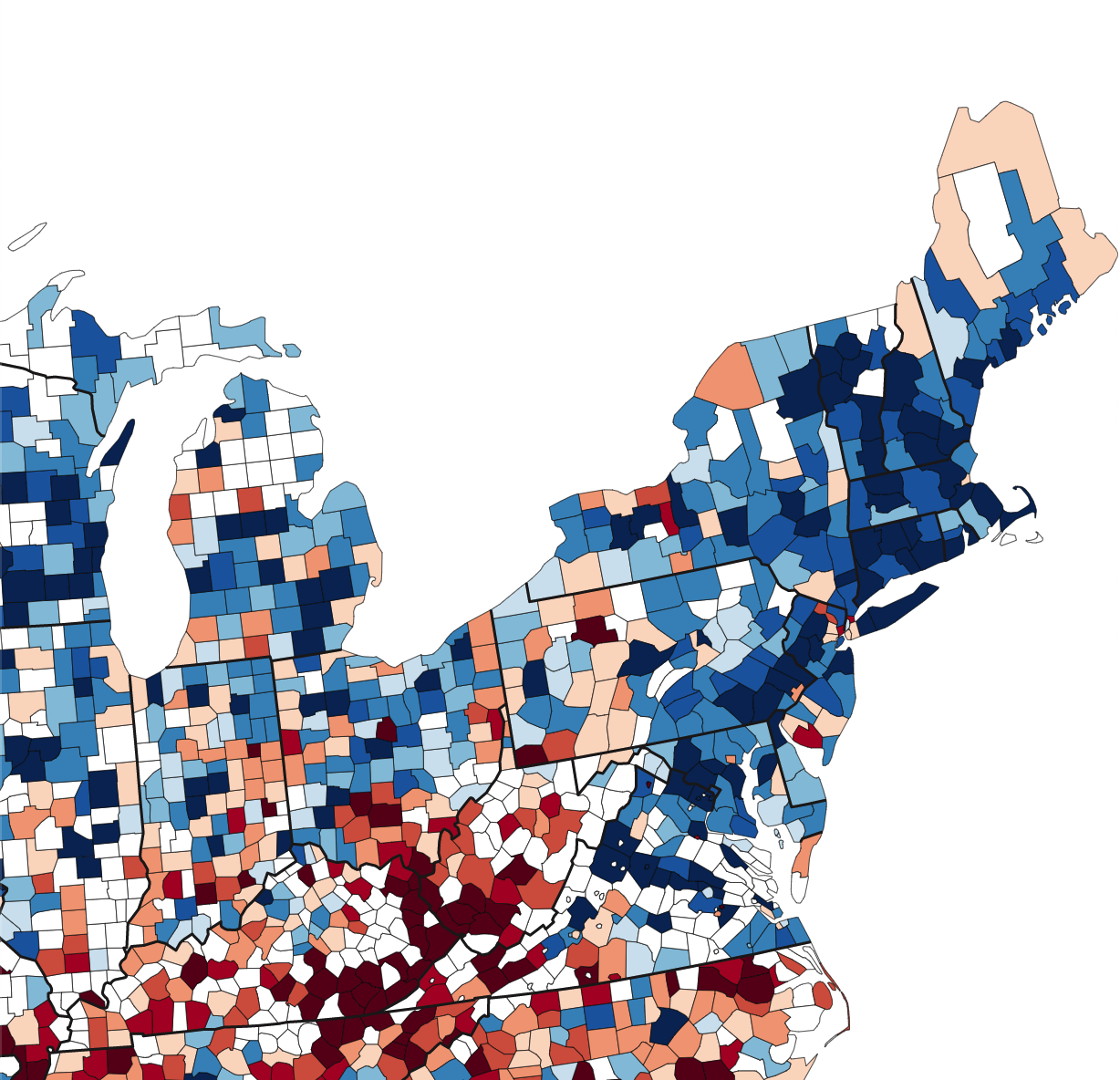}
  \caption{BRFSS Reported Poor/Fair Health}
  \label{fig:sfig1}
\end{subfigure}%
\begin{subfigure}{.5\textwidth}
  \centering
  \includegraphics[width=.6\linewidth]{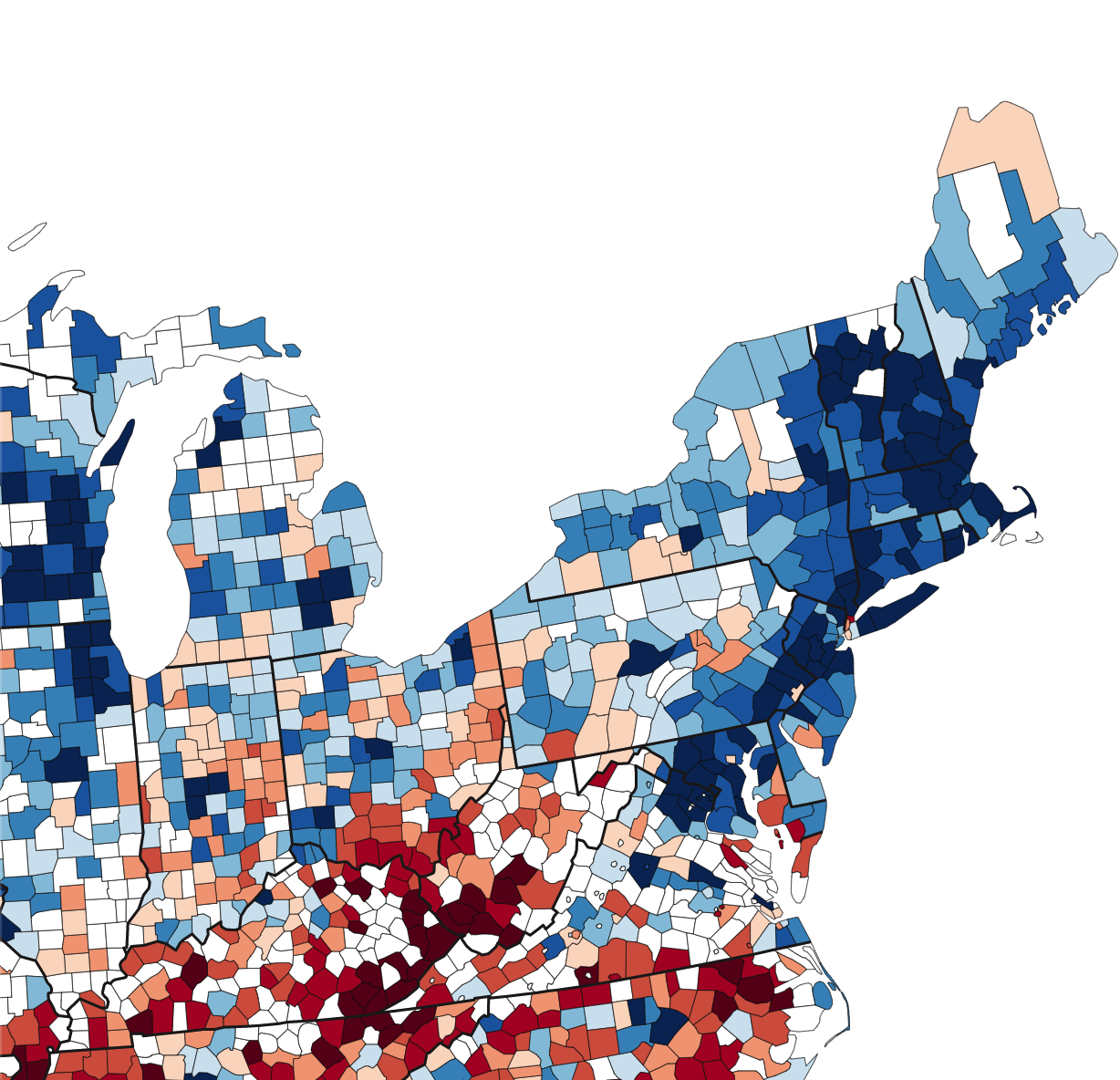}
  \caption{Twitter Predicted Poor/Fair Health}
  \label{fig:sfig2}
\end{subfigure}

\hspace{60mm}
\begin{subfigure}{.5\textwidth}

  \includegraphics[width=.6\linewidth]{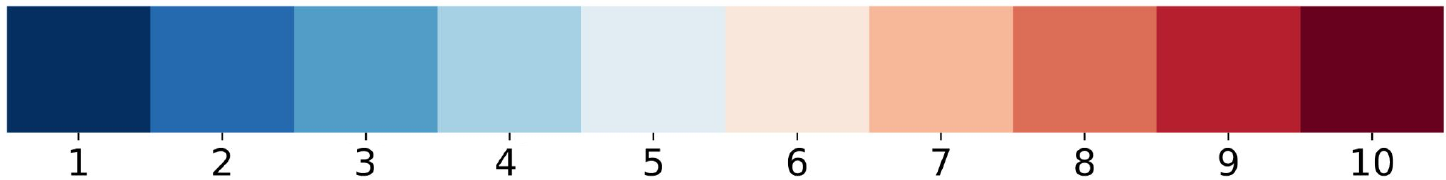}
\end{subfigure}
\caption{Maps of northeastern U.S. counties showing the deciles of adults living in Poor/Fair Health (a) as reported by the BRFSS and (b) as predicted from Twitter, using our recommended model in Table \ref{table:best model}. Out-of-sample Twitter predictions obtained though the cross validation process. Blue is less poor/fair health (i.e., better health); red is more poor/fair health (i.e., worse health), and white is unreliable self-report or Twitter data (i.e., missing data). }
\label{fig:county maps}
\end{figure*}

\paragraph{\textbf{Recommended System}} Due to the large number of tuning parameters evaluated above, we perform a backwards elimination on our correction variables to find the model with the highest accuracy and the smallest number of correction factors.
To make this selection process simpler, we evaluate our backwards selection using raking, as this method tended to outperform naive post-stratification, as well as estimator redistribution. 
The algorithm behaves as follows. First, we start with the maximum number of correction factors (age, gender, income, and education) and then perform a grid search over Adaptive Binning and Informed Smoothing parameters. We then choose a smaller set of correction factors and perform the same grid search, until we find a model with the smallest number of correction factors that gives us the best accuracy. This evaluation is performed across our four county-level outcomes, as opposed to an average of the four.
Table \ref{table:best model} shows predictive accuracies for each of our four outcomes using the above algorithm. 
Here we see a fairly consistent set of tuning parameters: a minimum bin threshold of 50, a smoothing constant $k=10$, and income as a correction factor. 
For two of the our outcomes, heart disease and poor/fair health, we see education as an additional correction factor.
The maximum percent increase occurs for life satisfaction (52.9\%) while heart disease has the smallest significant increase (4.44\%).
In the end, both models are very similar, with the only difference being the addition of education correction. 
Thus, in the end, we recommend using both income and education since, everything else being equal, we believe it is better to correct for more factors in order to make the model more ``fair".
In Figure \ref{fig:county maps} show county level heat maps of the true values of the Poor/Fair Health measure (as reported by the BRFSS) and Twitter predicted values (as determined by cross validation using our recommended system). 
Here we see our recommended system in Figure \ref{fig:county maps}(a) reasonably tracking the ground truth values in Figure \ref{fig:county maps}(b).

\subsection{Quantifying Bias Reductions}

Here we quantify the reduction in bias in our recommended system. For our two continuous variables (age and income) we take the absolute difference in means of the census bin percentages and our estimated demographics, normalized by the pooled standard deviation, and average across all counties (i.e., average absolute Cohen's D). We do this once for our ``out-of-the-box" socio-demographic estimates (to get a baseline bias) and again for the weighted estimates from our recommended model (i.e., correcting for income and education with a minimum bin threshold of 50 and smoothing $k=10$). For the two categorical variables (gender and education), we take the absolute difference in county percentage female and percentage with a bachelor's degree, and then average across all counties. 
This process is repeated for each of our four county level tasks. 

Figure \ref{fig:quantifying bias} shows the results of this experiment. Here we see a significant reduction in bias for age, income, and education across all four tasks. Gender, on the other hand, shows a slight increase in bias, though we note that this increase is not nearly as dramatic as the decreases seen across age, income, and education, nor have we attempted to fully correct for gender (as this model corrects for income and education only).
These results seem to match the county-level statistics reported in Table \ref{table:county means}, which shows a small difference difference in gender between the Census data and our Twitter sample. 
That is, we do not expect our methods to drastically address gender biases because gender is evenly distributed across counties, in both the Census data and our Twitter sample. 
Finally, we note that, while our recommended system only performs raking across income and education, we apply estimator redistribution across all four correction factors (age, gender, income, and education). Thus, we might expect our final bias measures to correct for age despite the fact our post-stratification process does not consider this variable.

\section{Conclusion}
\label{section:discussion}
Selection bias --- producing measurements over a population sample that differs from the target population --- is a frequent criticism of automatic social media-based population predictions~\cite{hoover2018big,wang2019demographic,shah2020predictive}.
While post-stratification techniques are frequently used to address selection bias in opinion polling or social sciences, we found ``out of the box'' methods generally resulted in worse performance, as compared to no correction, for the task of predicting population (i.e., U.S. county) health and well-being statistics from social media language. 
We discovered two reasons for this lack of benefit: (1) \emph{estimating} sample user demographics from predictive models (as opposed to having self-reported demographics of the Twitter users) introduces additional biases when compared to known distributions and (2) sparse or underpowered data for estimating the observed community demographic distributions. 
To the best of our knowledge, neither of these issues has been previously investigated for improving post-stratification. 
In fact, few works have even evaluated commonly used selection bias mitigation techniques for predictive tasks~\cite{wang2019demographic,culotta2014reducing}, likely because such techniques are traditionally applied without access to ground truth validation data (e.g., in most opinion polls). 

We proposed \name which includes several techniques to address challenges in selection bias correction for predicting population statistics and evaluating their efficacy.
First, we found that using \textit{estimator redistribution} to counter shrinkage bias of estimated demographics provided modest benefits. 
Then, we explored two techniques for addressing sparse bin issues: \textit{adaptive binning} and \textit{informed smoothing}, finding both provided a substantial benefit and resulted in an overall improvement to the predictive models, yielding state-of-the-art results (a 52.9\% increase in variance explained for life satisfaction).
Many approaches for addressing demographic biases in AI try to correct them without sacrificing accuracy~\cite{gonen-goldberg-2019-lipstick-pig}. 
In the case of selection bias, we believe we have shown that properly correcting bias can yield substantial benefits.

\bibliography{references}

\end{document}


\maketitle

\section{Cross Correlations}
\label{section:correlations}

The correlations between our four health outcomes and four socio-demographic variables are presented in Table \ref{table:correlations}. When correcting for selection biases it is not always clear which biases exist in your sample. While this paper was limited to age, gender, income, and education, there exist many other variables one could correct for. 
To this end, we present the correlations between all variables used in this paper, in order to interpret our results --- given these relationships, would we expect correcting for certain socio-demographic variables to increase predictive performance on a specific outcome? The correlations in Table \ref{table:correlations} show that both income and education are highly associated with all of our outcome variables. On the other hand, age and gender are not, with the exception of suicide. Given the size of the income correlations we might expect correcting for income to give us the biggest benefit.

\begin{table}[htpb!]\centering
\resizebox{\columnwidth}{!}{\begin{tabular}{lccc|cccc}\toprule
\multicolumn{1}{c}{}  & Suicide  & \begin{tabular}[c]{@{}c@{}}Life\\ Satisfaction\end{tabular} & \begin{tabular}[c]{@{}c@{}}Fair/Poor\\ Health\end{tabular} &  Income & Education & \begin{tabular}[c]{@{}c@{}}Percent\\ Female\end{tabular} & \begin{tabular}[c]{@{}c@{}}Median\\ Age\end{tabular} \\ \hline
Heart Disease  & .18 & -.34 & .60 & -.58 & -.59 & .09 & .03 \\
Suicide  & - & -.04 & .20 & -.31 & -.34 & -.15 & .30 \\
Life Satisfaction  &   & - & -.35 & .37 & .39 & -.06 & -.05 \\
Fair/Poor Health   &  &  & - & -.65 & -.62 & .06 & .02 \\ \bottomrule
\end{tabular}}
\caption{Correlations between county level health and socio-demographic variables}
\label{table:correlations}
\end{table}
\section{Method Examples}
\label{section:examples}

Here we give brief examples of our more complex methods to aid in understanding.

\paragraph{Estimator Redistribution} Here we redistribute our estimated socio-demographics at the national level (i.e., across all counties) to match the national distribution reported by PEW. 
(See \textbf{Data} for a description of the PEW data.)
The national percentage of people on Twitter between $\min_h^{(t)}=18$ and $\max_h^{(t)}=29$ is 51.1\% (as reported by PEW's Social Media update~\cite{duggan2015social,greenwood2016social}; averaged across 2013-2016).  
We then start with our minimum predicted age in our sample $\min_h^{(s)}=13$ and then find the age $\max_h^{(s)}$ such that the percentage of Twitter users in our sample between 13 and $\max_h^{(s)}$ equals 51.1\%.
We then adjust all age estimates with that bin using Equation 6.
Next, we set $\min_h^{(t)}=30$ and $\max_h^{(t)}=49$ and note that the PEW reported national percentage of people on Twitter in this age bin is 31.8\%.
We then set $\min_h^{(s)}$ equal to $\max^{(s)}$ from the previous iteration.
Finally, we find the age $\max_h^{(s)}$ such that the percentage of Twitter users in our sample between $\min_h^{(s)}$ and $\max_h^{(s)}$ equals 31.8\%.
This process is repeated for all bins.

\paragraph{Adaptive Binning} In adaptive binning we set a minimum number of observations per bin and collapse all bins with the smallest adjacent bin if they do not meet this threshold. 
This is repeated until all bins meet the threshold or we have a single bin.
For example, in a given county we have at most 11 age bins. 
We start with the bin with the smallest number of Twitter users and see if this number meets our minimum threshold. 
In this example lets define our minimum bin threshold as 50 and assume the age bin with the smallest number of Twitter users mapped to our county is 45-49 years old.
We check if there are at least 50 users who are between 45 and 49 years old.
If not, we combine this bin with the smallest, adjacent bin (either 40-44 or 50-54). 
Assume that the bin 40-44 has less users than 50-54.
We combine then combine the bins 40-44 and 45-49, resulting in the bin 40-49, and discard the bins 40-44 and 45-49.
We then start the process over: identify the smallest bin and repeat the above steps until all bins meet our threshold or we have a single bin.

\section{Additional Experiments}
\label{section:additional experiments}

\paragraph{Add One Smoothing} We examine the effects of ``add one" smoothing in Table \ref{table:uninformed smoothnig}. Here we add 1 to each socio-demographic bin. We see that the prediction accuracies are comparable to binning, with a minimum bin size of 1, and smoothing with $k=1$, though this method fails to increase performance over baseline. 

\begin{table}[ht]\centering
\resizebox{\columnwidth}{!}{
\begin{tabular}{lccc}\toprule
 \textbf{} & Post-stratification & \begin{tabular}[c]{@{}c@{}}Naive \\ Post-Statification\end{tabular} & Raking \\ \hline
Age & .587 & - & - \\
Gender & .641 & - & - \\
Income & .617 & - & - \\
Education & .644 & - & - \\
Age + Gender & - & .590 & .607 \\
Income + Education & - & .629 & .630 \\
Age + Gen. + Inc. + Edu. & - & .640 & .628 \\
\bottomrule
\end{tabular}}
\caption{Evaluation of ``add one" Smoothing. Results are comparable to adaptive binning with bin size = 1 and informed smoothing with k = 1, but no increase above baseline.}
\label{table:uninformed smoothnig}
\end{table}

\bibliography{references}